 \documentclass[structabstract]{aa}  
\usepackage{graphicx}
\usepackage{txfonts}
\usepackage{natbib}
\bibpunct{(}{)}{;}{a}{}{,}

\newcommand\days[2]{#1\hskip1.5pt.\hskip-4pt$^{\rm{d}}$#2}

\begin{document}

   \title{Absolute dimensions of solar-type eclipsing binaries.\thanks{Based on observations carried out at the Str\"{o}mgren Automatic Telescope (SAT) at ESO, La Silla, and the Mercator Telescope, operated on the island of La Palma by the Flemish Community, at the Spanish Observatorio del Roque de los Muchachos of the Instituto de Astrofísica de Canarias.} \thanks{Table X (light curves of EF Aqr) is only available in electronic form at the CDS via anonymous ftp to cdsarc.u-strasbg.fr (130.79.128.5) or via http://cdsweb.u-strasbg.fr/cgi-bin/qcat?J/A+A/}}

   \subtitle{EF Aquarii: a G0 test for stellar evolution models.}

   \author{J. Vos
          \inst{1,2}
          \and
          J.V. Clausen
	  \inst{2}
	  \and
	  U.G. J{\o}rgensen
	  \inst{2,3}
	  \and
	  R.H. \O{}stensen
	  \inst{1}
          \and
	  A. Claret
	  \inst{4}
	  \and
	  M. Hillen
	  \inst{1}
	  \and
	  K. Exter
	  \inst{1}
          }

   \institute{Instituut voor Sterrenkunde, K.U.Leuven,Celestijnenlaan 200D BUS 2401, 3001 Leuven, Belgium\\
             \email{jorisv@ster.kuleuven.be}
             \and
             Niels Bohr Institute, Copenhagen University, Juliane Maries Vej 30, 2100 Copenhagen \O{}, Denmark
             \and
             Centre for Star and Planet Formation, Geological Museum, {\O}ster Voldgade 5, 1350 Copenhagen K, Denmark
             \and
	     Instituto de Astrof\'{\i}sica de Andaluc\'{\i}a, CSIC, Apartado 3004, 18080 Granada, Spain
             }

   \date{Received \today; accepted ???}

 
  \abstract
   {Recent studies have shown that stellar chromospheric activity, and its effect on convective energy transport in the envelope, is most likely the cause of significant radius and temperature discrepancies between theoretical evolution models and observations. Accurate mass, radius, and abundance determinations from solar-type binaries exhibiting various levels of activity are needed for a better insight into the structure and evolution of these stars.}
   {We aim to determine absolute dimensions and abundances for the solar-type detached eclipsing binary EF Aqr, and to perform a detailed comparison with  results from recent stellar evolutionary models.}
   {$uvby$ light curves and $uvby\beta$ standard photometry were obtained with the Str\"{o}mgren Automatic Telescope. The broadening function formalism was applied on spectra observed with HERMES at the Mercator telescope in La Palma, to obtain radial velocity curves. State-of-the-art methods were applied for the photometric and spectroscopic analyses.}
   {Masses and radii with a precision of 0.6\% and 1.0\% respectively have been established for both components of EF Aqr. The active 0.956 $M_{\odot}$ secondary shows star spots and strong Ca II H and K emission lines. The 1.224 $M_{\odot}$ primary shows signs of activity as well, but at a lower level. An [Fe/H] abundance of 0.00$\pm$0.10 is derived with similar abundances for Si, Ca, Sc, Ti, V, Cr, Co, and Ni. Solar calibrated evolutionary models such as Yonsei-Yale, Victoria-Regina and BaSTI isochrones and evolutionary tracks are unable to reproduce EF Aqr, especially for the secondary, which is 9\% larger and 400 K cooler than predicted. Models adopting significantly lower mixing length parameters $l/H_p$ remove these discrepancies, as seen in other solar type binaries. For the observed metallicity, Granada models with a mixing length of $l/H_p=1.30$ (primary) and 1.05 (secondary) reproduce both components at a common age of 1.5$\pm$0.6 Gyr.}
   {Observations of EF Aqr suggests that magnetic activity, and its effect on envelope convection, is likely to be the cause of discrepancies in both radius and temperature, which can be removed by adjusting the mixing length parameter of the models downwards.}


   \maketitle
%

\defcitealias{Clausen01}{CHO01}
\defcitealias{Clausen08}{CTB08}
\defcitealias{Clausen09}{CBC09}

\section{Introduction}
As recent studies suggest, current stellar evolutionary models scaled to the Sun, are unable to predict the temperature and radii of many active close binary components with masses comparable to, or lower than the Sun.  These models predict too high temperatures and radii that are up to 10\% lower than observed, while the observations have errors as small as about 1\%, see e.g. \citet[hereafter CBC09]{Clausen09} and references therein. 
Dynamo--generated magnetic fields can suppres convection and produce starspots, and are suggested as a cause for the discrepancies in radii and temperature, see e.g. \citet{Ribas08} for an overview. \citet{Clausen99} and \citet{Torres06} have shown that by reducing the mixing length parameter when calculating the envelope convection, the model fits can be improved. \citet{Clausen09} found that solar-type eclipsing binaries can (with one exception) be divided in to two groups: slowly rotating inactive binaries, which can be fitted reasonably well by models with a solar calibrated mixing length, and rapidly rotating systems, which exhibit intrinsic variations and show signs for increased magnetic activity, which cannot be reproduced by solar-calibrated models. These authors suggested that the evolution of the mixing length parameter with mass and activity is influenced by two different effects. There is a slight decrease in mixing length with increasing temperature and mass for inactive main sequence stars, and a strong decrease in the mixing length parameter for active stars compared to their inactive counterparts of similar mass. However, more binaries are needed to develop a calibration for the evolution of mixing length with physical parameters and level of activity.

This paper is part of our ongoing study of several binary systems with solar--type components, with either identical or different levels of chromospheric activity \citep[hereafter CHO01]{Clausen01}. In this paper the results of a complete analysis of the G0 system EF Aqr (which has two components of different and relatively high activity) are presented.

\section{EF Aqr}
EF Aqr ($m_V$ = 9.88, spectral type G0, \days{2}{85}, also known as BD$-$7$^{\circ}$5908, BV207, and as HD217512) is a solar--type eclipsing binary that has not been studied in detail before. It was first discovered at the Bamberg observatory by \citet{Strohmeier58}. In 1959 several times of minima were published by \citet{Nikulina59}. Their studies do not agree well with the times of minima obtained for this paper, and were not used. In 2006 new elements were published by \citet{Otero06} based on a combination of data found in different catalogues. Both the published epoch and period differ from our results presented in Table \ref{tb_times_of_minima} and Table \ref{tb_absolute_dimensions}, most likely because of the relatively low quality of the data Otero and collaborators were able to collect. In 2007, EF Aqr was included in the variable star one shot project, and one HARPS spectrum was taken \citep{Dall07}. It was analysed as a single--lined spectrum, resulting in an effective temperature of 6110$\pm$187 K, a $\log{g}$ of $4.36 \pm 0.59$ and a metallicity of [Fe/H]$ = 0.24 \pm 0.18$.

\section{Photometry}
Below, we present the new photometric material for EF Aqr and refer to \citetalias{Clausen01} for more details on the observation and reduction procedures, and determination of the times of minima.

\subsection{Light curves}
The differential $uvby$ light curves of EF Aqr where obtained at the Str\"{o}mgren Automatic Telescope (SAT) at ESO, La Silla and its six-channel $uvby\beta$ photometer, on 80 nights between October 2001 and December 2008 (JD 2452185 to 2454821). The observations were made through an 18 arcsec diameter circular diaphragm. The light curves contain 764 points per band, and most orbital phases are covered at least twice. HD 217376, HD 217877, and HD 218730 -- all within a few degrees of EF Aqr on the sky -- were used as comparison stars and were all found to be constant within a few mmag; see Table \ref{tb_photometric_data}. The light curves are calculated relative to HD 217877, but all comparison star observations were used, shifting them first to the same light level. The average accuracy per light curve point is 4--5 mmag ($vby$) and 5--6 mmag ($u$). The light curves (\textbf{Table X}) are available in electronic form at CDS. 

\begin{figure*}[!t]
\centering
\includegraphics{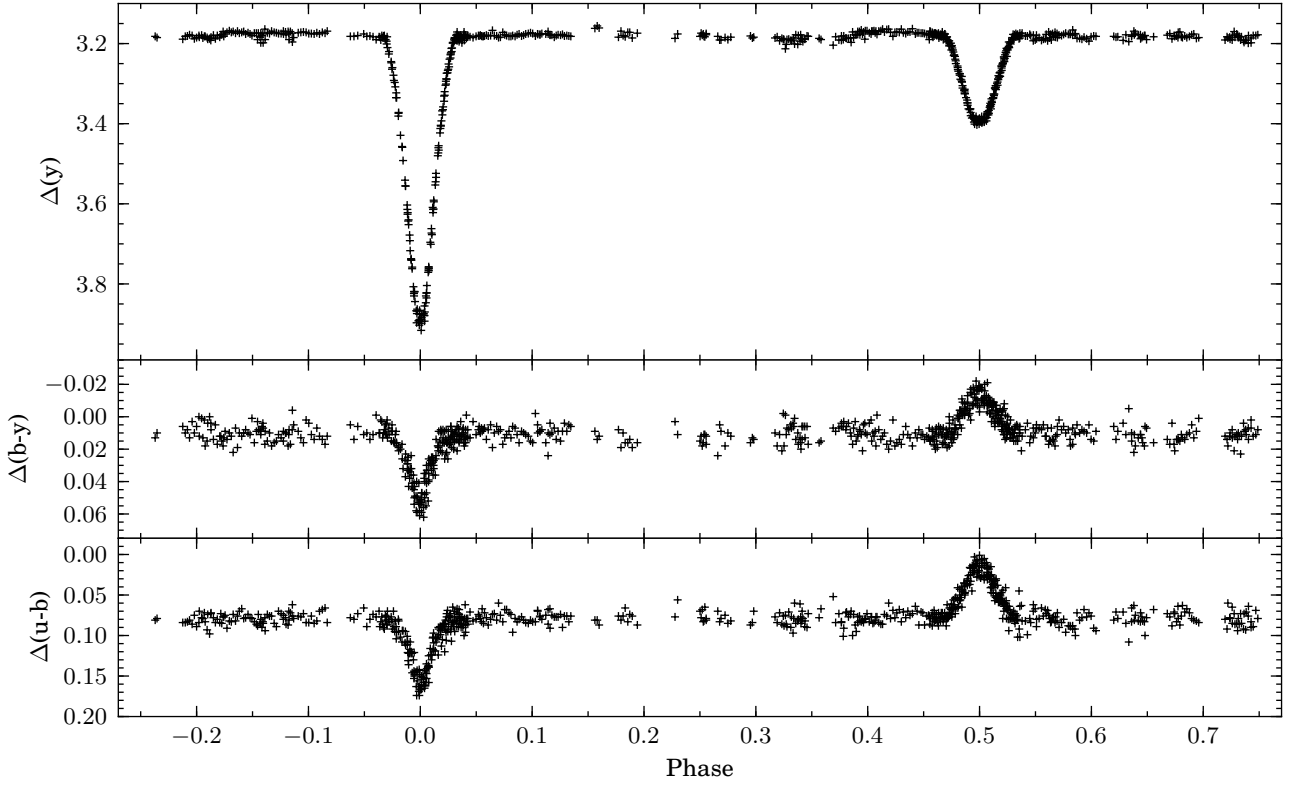}
\caption{$y$ light curve and $b-y$ and $u-b$ colour curves (instrumental system) for EF Aqr.}\label{fig_light_curve}%
\end{figure*}

As seen in Fig. \ref{fig_light_curve}, EF Aqr is a slightly active, detached system with quite different eclipse depths of about 0.7 mag and 0.2 mag (in $y$). The secondary eclipse occurs at phase 0.5 and the duration of the two eclipses is identical, meaning that the orbit is circular. 

\subsection{Times of minima and ephemeris}
Four of each of the primary and secondary minima have been determined from the $uvby$ light curves, and  are listed in Table \ref{tb_times_of_minima}. 

\begin{table}
\caption{Times of primary (P) and secondary (S) minima of EF Aqr determined from the $uvby$ observations. }
\label{tb_times_of_minima}
\centering                         
\begin{tabular}{cccc}      
\hline\hline               
HJD & rms & Type & O$-$C\tablefootmark{a}\\
$-2400000$ & & & \\
\hline
52225.55854	& 0.00014	& P	& 0.00014\\
52556.57275	& 0.00010	& P	& $-$0.00001\\
52576.54774	& 0.00011	& P	& $-$0.00003\\
52764.88340	& 0.00100	& P	& $-$0.00012\\
52563.70652	& 0.00026	& S	& $-$0.00017\\
52794.84594	& 0.00014	& S	& $-$0.00009\\
52917.54967	& 0.00029	& S	& 0.00004\\
53676.59990	& 0.00019	& S	& 0.00011\\
\hline                                   
\end{tabular}
\tablefoot{\tablefoottext{a}{O-C values are calculated for the ephemeris given in Eq. \ref{eq_ephemeris}, assuming a circular orbit.}}
\end{table}

By adopting a weighted least--squares fit to the times of primary and secondary minima (i.e. assuming a circular orbit as argued for above), we derive the linear ephemeris given as        
\begin{equation}
 \mathrm{Min I} = 2452556.57276(6) + $\days{2}{85357206(44)}$ \times E\label{eq_ephemeris}
\end{equation}

\subsection{Standard photometry}
Standard $uvby\beta$ indices for EF Aqr and the comparison stars, observed and derived as described by \citetalias{Clausen01}, are presented in Table \ref{tb_photometric_data}. The indices are based on many observations and have a high precision. For comparison we included published photometry from other sources as well. In general, the agreement is good, but in some cases differences are larger than the quoted errors. We have used only the new results from our own work for the analysis of EF Aqr presented here.

\begin{table*}[!t]
\caption{Photometric data for EF Aqr and the comparison stars.}             
\label{tb_photometric_data}      
\centering          
\begin{tabular}{lllrrrrrrrrrrrr}    
\hline\hline
  Object	&   Spec. Type	&	Ref.	&	V	&   $\sigma$	&	$b-y$	&   $\sigma$	&	$m_1$	&   $\sigma$	&	$c_1$	&    $\sigma$	&   N($uvby$)	&	$\beta$	&   $\sigma$	&     N($\beta$)\\\hline	
EF Aqr		&	G0\tablefootmark{a}	&	V11	&	9.885	&	9	&	0.382	&	5	&	0.191	&	8	&	0.374	&	10	&	280	&	2.621	&	9	&	8	\\
		&		&	V11	&	10.092	&	9	&	0.356	&	4	&	0.178	&	6	&	0.386	&	11	&	22	&		&		&		\\\hline
HD217376	&    F3 IV\tablefootmark{b}	&	V11	&	6.807	&	5	&	0.288	&	4	&	0.149	&	7	&	0.478	&	6	&	192	&	2.648	&	7	&	4	\\
		&		&	O83	&	6.805	&	6	&	0.292	&	4	&	0.141	&	6	&	0.472	&	6	&	1	&		&		&		\\
		&		&	O94	&	6.810	&	4	&	0.289	&	3	&	0.141	&	4	&	0.480	&	5	&	1	&	2.645	&	6	&	1	\\\hline
HD217877	&    G2 V\tablefootmark{b}	&	V11	&	6.704	&	4	&	0.371	&	4	&	0.174	&	7	&	0.346	&	7	&	215	&	2.608	&	8	&	4	\\
=HR8772		&		&	P69	&	6.65	&		&	0.364	&		&	0.181	&		&	0.319	&		&	2	&		&		&		\\
		&		&	GO76	&	6.702	&	6	&	0.367	&	4	&	0.175	&	6	&	0.350	&	22	&	3	&		&		&		\\
		&		&	GO77	&		&		&		&		&		&		&		&		&		&	2.601	&	8	&	5	\\
		&		&	AJ85	&		&		&		&		&		&		&		&		&		&	2.602	&	2	&	11	\\\hline
HD218730	&    G1/2 V\tablefootmark{b}	&	V11	&	7.333	&	5	&	0.385	&	5	&	0.198	&	7	&	0.361	&	8	&	180	&	2.608	&	7	&	4	\\
		&		&	O83	&	7.334	&	6	&	0.376	&	4	&	0.207	&	6	&	0.346	&	6	&	1	&		&		&		\\
		&		&	O94	&	7.331	&	4	&	0.384	&	3	&	0.195	&	4	&	0.357	&	5	&	1	&		&		&		\\
\hline                  
\end{tabular}
\tablefoot{
For EF Aqr, the $uvby\beta$ information by V11 is the mean value near phases 0.25 and 0.75 (first line), and at total secondary eclipse (second line). \\
N is the total number of observations used to form the mean values, and $\sigma$ is the rms error (per observation) in mmag.\\
\tablefoottext{a}{Henry Draper Catalogue}; \tablefoottext{b}{\citet{Houk99}}}
\tablebib{
AJ85 = \cite{Andersen85}; GO76 = \cite{Gronbech76}; GO77 = \cite{Gronbech77}; O83 = \cite{Olsen83}; O94 = ($uvby$): \cite{Olsen94}, ($\beta$): Olsen unpublished; P69 = \cite{Perry69}; V11 = this paper }
\end{table*}

\subsection{Photometric elements}
Since EF Aqr is a well--detached system, JKTEBOP \citep{Southworth04_a, Southworth04_b} has been used to determine the photometric elements of the $uvby$ light curves. JKTEBOP is based on the Nelson-Davis-Etzel binary model \citep{Nelson72, Etzel81, Martynov73}, which represents deformed stars as biaxial ellipsoids and applies a simple bolometric reflection model. More details on the code and the general approach used here is explained in \citet[hereafter CTB08]{Clausen08}. Hereafter the following symbols are used (with subscript $s$ and $p$ referring to respectively the secondary and the primary component of the system): $i$ is the orbital inclination; $e$ the eccentricity of the orbit; $\omega$ the longitude of periastron; $r$ the stellar radius in units of the semi-major axis; $k=r_s/r_p$; $u$ is the linear limb darkening coefficient; $y$ the gravity darkening coefficient; $J$ the central surface brightness; $L$ is the luminosity and $T_{\mathrm{eff}}$ the effective temperature. 

The mass ratio between the components was kept at the spectroscopic value ($q=0.761$); see section \ref{spectroscopic_elements}. Linear limb darkening coefficients from \citet{Claret00} were adopted. Extensive tests with nonlinear limb darkening laws and Van Hamme coefficients \citep{VanHamme93} were conducted, but showed little deviation from the results presented here. The linear limb darkening law with Claret coefficients was adopted because it resulted in the most consistent results between the four bands. Gravity darkening coefficients from \citet{Bloemen11} for convective atmospheres were adopted. A circular orbit is assumed, fixing $e\cos{\omega} = e\sin{\omega} = 0$. We Justify this assumption in section \ref{spectroscopic_elements}. 

The results for EF Aqr are presented in Table \ref{tb_photometric_solutions}. The results from the four bands agree well. The errors quoted in this table are formal errors determined from the iterative least--squares solution procedure. Changing to VanHamme limb darkening coefficients, or adopting a nonlinear limb darkening law (quadratic, logarithmic and square root) changes the resulting inclinations by about $0.1^{\circ}$, and the radii by $0.005$.

\begin{table}
\caption{Photometric solutions for EF Aqr from the JKTEBOP code.}
\label{tb_photometric_solutions}
\centering
\begin{tabular}{lrrrr}
\hline\hline
	&	$u$	&	$v$	&	$b$	&	$y$	\\\hline
$i$ ($^{\circ}$)	&	88.40	&	88.46	&	88.44	&	88.49	\\
	&	$\pm$  7	&	$\pm$ 6	&	$\pm$ 5	&	$\pm$ 5	\\
$r_p$	&	0.1224	&	0.1213	&	0.1218	&	0.1212	\\
$r_s$	&	0.0875	&	0.0870	&	0.0873	&	0.0876	\\
$k$	&	0.7152	&	0.7170	&	0.7168	&	0.7159	\\
	&	$\pm$ 25	&	$\pm$ 19	&	$\pm$ 17	&	$\pm$ 15	\\
$r_p+r_s$	&	0.2099	&	0.2084	&	0.2091	&	0.2080	\\
	&	$\pm$ 6	&	$\pm$ 5	&	$\pm$ 5	&	$\pm$ 5	\\
$u_p$	&	0.8107	&	0.7673	&	0.7369	&	0.6553	\\
$u_s$	&	0.9223	&	0.8796	&	0.8323	&	0.7478	\\
$y_p$	&	0.6527	&	0.5648	&	0.4465	&	0.3791	\\
$y_s$	&	1.2437	&	0.8670	&	0.6904	&	0.5580	\\
$J_s/J_p$	&	0.2438	&	0.2851	&	0.3729	&	0.4251	\\
	&	$\pm$ 27	&	$\pm$ 21	&	$\pm$ 19	&	$\pm$ 19	\\
$L_s/L_p$	&	0.1184	&	0.1410	&	0.1834	&	0.2092	\\
$\sigma$ (mmag)	&	10.9	&	8.3	&	7.5	&	7.3	\\

\hline
\end{tabular}
\tablefoot{Linear limb darkening coefficients from \citet{Claret00} were adopted together with gravity darkening coefficients from \citet{Bloemen11}. A mass ratio of $q=0.761$ from the spectroscopy was assumed, the orbit was assumed to be circular, and phase shift and magnitude normalisation were included as free parameters. \\
The quoted errors are formal errors from the iterative least--squares solution procedure.}
\end{table}

The limb darkening coefficients are based on the effective temperature of both components. The effective temperature was calculated from a weighted mean of three colour-metallicity-temperature calibrations, the spectroscopic temperature (see Section \ref{chemical_abundances}), and for the secondary the flux scale method of \citet{Popper80} is included as well (see Section \ref{absolute_dimensions}). To obtain the dereddened colour indices, a reddening of $E(b-y) = 0.018\pm0.011$ was determined with the intrinsic colour calibration of \citet{Olsen88}.

The lightcurves show clear indications for star spots. PHOEBE \citep{Prsa06}, a binary modelling code built on the Wilson-Devinney code \citep{Wilson08}, was used to analyse separate sections of the light curves while including star spots in the model. Sections of different observation periods ranging from five months to two years were tested, but PHOEBE was not able to fit the effect of the star spots.

\begin{table}
\caption{Adopted photometric elements for EF Aqr.}
\label{tb_adopted_photometric_elements}
\centering
\begin{tabular}{lrrrr}
\hline\hline
&	\multicolumn{1}{l}{$i$}		&	\multicolumn{2}{l}{88.454 $\pm$ 0.074}	&	\\
&	\multicolumn{1}{l}{$e$\tablefootmark{a}}	&	\multicolumn{2}{l}{0.0}		&	\\
&	\multicolumn{1}{l}{$\omega$\tablefootmark{a}}	&	\multicolumn{2}{l}{90.0}	&	\\
&	\multicolumn{1}{l}{$r_p$}	&	\multicolumn{2}{l}{0.1216 $\pm$ 0.0012}	&	\\
&	\multicolumn{1}{l}{$r_s$}	&	\multicolumn{2}{l}{0.0871 $\pm$ 0.0003}	&	\\
&	\multicolumn{1}{l}{$r_p + r_s$}	&	\multicolumn{2}{l}{0.2098 $\pm$ 0.0013}	&	\\
&	\multicolumn{1}{l}{$k$}		&	\multicolumn{2}{l}{0.7164 $\pm$ 0.0063}	&	\\
		&	$u$	&	$v$	&	$b$	&	$y$	\\
$J_s/J_p$	&	0.244	&	0.285	&	0.373	&	0.425	\\
		& 	$\pm$9	&	$\pm$9	&	$\pm$12	&	$\pm$13	\\
$L_s/L_p$	&	0.1188	&	0.1413	&	0.1837	&	0.2096	\\
		&	$\pm$19	&	$\pm$20	&	$\pm$17	&	$\pm$17	\\	
\hline
\end{tabular}
\tablefoot{\tablefoottext{a}{adopted parameters}\\
The individual flux and luminosity ratios are based on the mean stellar and orbital parameters.}
\end{table}

The adopted photometric elements listed in Table \ref{tb_adopted_photometric_elements} are the weighted mean values of the JKTEBOP solutions in the four bands. Realistic errors were determined from Monte Carlo error simulations with 5000 iterations in each band and a comparison between the solutions in each band. The uncertainty of the limb darkening values based on the accuracy of the effective temperature was taken into account as well. The accurate light curves allow the relative radii to be determined with a precision higher than 1\%. The obtained luminosity ratios correspond very well to those derived from the depths of secondary eclipse. 

\section{Spectroscopy}
For the radial velocity and abundance determinations of EF Aqr, 17 high--resolution spectra were obtained with the HERMES spectrograph (R = 85000, 55 orders, 3770-9000 \AA, \citet{Raskin11}), at the 1.2m Mercator telescope, Roque de los Muchachos Observatory, La Palma. The observations were obtained in October and November 2010 (JD 2455509-2455545). HERMES was used in high--resolution mode, and Th-Ar-Ne frames were obtained just before or just after the observations to reduce the instrumental drift. The exposure times were adjusted to obtain a signal to noise ratio (S/N) around 100 in V.  For the basic reduction of the spectra the HERMES pipeline version 3.0 was used together with dedicated IDL programmes to further clean and normalise the spectra. For the radial velocity determinations only spectra from orders 13-48 (4016-6796 \AA) were used, because the S/N in the other orders was too low.

\subsection{Radial velocities}
To obtain the radial velocities from the spectra, the broadening function (BF) formalism \citep{Rucinski99, Rucinski02,Rucinski04} was used on 16 spectra. The last spectrum, which was observed just after secondary eclipse, was not included in the radial velocity determinations. Template spectra calculated for the primary and secondary components were used in the analysis. The synthetic template spectra were computed with the \textit{bssynth} tool (Bruntt, private communication), which applies the SYNTH software \citep{Valenti96} and modified ATLAS9 models \citep{Heiter02}. Regardless of the significant temperature difference between the components, both templates give compatible results. But the radial velocities obtained with the primary template correspond slightly better to the final orbital solution. The broadening functions were calculated separately for the different orders, and then combined by using a weighted average based on the S/N level and the wavelength range of each order. 

The radial velocities were obtained from the broadening functions by fitting an analytical BF to the observed BF \citep{Kaluzny06}, which was derived by assuming two perfectly spherical rigid rotators. The assumptions made in the derivation of the analytical BF are valid for EF Aqr because both components show little or no deformation. Fitting Gaussians to the BF does not result in values significantly different from those presented in Tab.\,5.

The obtained radial velocities are given in Table \ref{tb_radial_velocities}. By examining the distribution of the radial velocities calculated separately for each order, the uncertainties of the radial velocities are estimated at 0.35 km s$^{-1}$ for the primary and 1.31 km s$^{-1}$ for the secondary.

\begin{table}
\centering
\caption{The radial velocities of EF Aqr together with the residuals of the final spectroscopic orbit given in Table \ref{tb_spectroscopic_elements}. Supscripts $p$ and $s$ denote determinations based on templates of the primary and the secondary.}
\label{tb_radial_velocities}
\begin{tabular}{ccrrrr}
\hline\hline
BJD  &  Phase  &  \multicolumn{1}{r}{RV$_p$}  &  \multicolumn{1}{r}{RV$_s$}  &  \multicolumn{1}{r}{(O$-$C)$_p$}  &  \multicolumn{1}{r}{(O$-$C)$_s$}  \\
$-$2400000 &	& \multicolumn{1}{r}{km s$^{-1}$} & \multicolumn{1}{r}{km s$^{-1}$} & \multicolumn{1}{r}{km s$^{-1}$} & \multicolumn{1}{r}{km s$^{-1}$}	\\
 \hline
55509.34736	&  	0.764	&  	79.69	  &  	$-$114.33	  &  $-$0.08	& $-$0.03 \\
55509.38693	&  	0.778	&  	78.74	  &  	$-$113.24	  &  $-$0.06	& $-0$.22 \\
55510.33598	&  	0.111	&  	-58.49	  &  	67.06	  	  &  $-$0.45	&  0.16 \\
55510.49097	&  	0.165	&  	-76.83	  &  	91.89	  	  &  $-$0.28	&  0.65 \\
55512.40508	&  	0.836	&  	67.96	  &  	$-$98.88	  &  $-$0.20	&  0.15 \\
55512.49166	&  	0.866	&  	58.41	  &  	$-$86.06	  &  $-$0.25	&  0.48 \\
55513.40673	&  	0.187	&  	-81.85	  &  	98.23	  	  &  $-$0.14	&  0.20 \\
55513.50138	&  	0.220	&  	-86.73	  &  	104.43	  	  &  0.03	& $-$0.24 \\
55514.51857	&  	0.577	&  	35.21	  &  	$-$55.25	  &  0.35	&  0.00	\\
55516.43170	&  	0.247	&  	-88.10	  &  	106.77	  	  &  0.13	&  0.17 \\
55517.32517	&  	0.560	&  	27.23	  &  	$-$44.85	  &  0.28	& $-$0.01	\\
55517.50485	&  	0.623	&  	55.03	  &  	$-$81.42	  &  0.32	& $-$0.07	\\
55518.38558	&  	0.932	&  	30.76	  &  	$-$50.13	  &  $-$0.22	&  0.02 \\
55545.38178	&  	0.392	&  	-56.43	  &  	64.66	  	  &  0.41	& $-$0.67	\\
55545.39625	&  	0.397	&  	-54.36	  &  	61.59	  	  &  0.37	& $-$0.96	\\
55545.40725	&  	0.401	&  	-52.80	  &  	59.65	  	  &  0.28	& $-$0.74	\\
\hline
\end{tabular}
\end{table}

\subsection{Spectroscopic elements}\label{spectroscopic_elements}
The spectroscopic orbits were calculated using the method of Lehman-Filh\'{e}s implemented in the SBOP programme \citep{Etzel81, Etzel04}, which is a modified and extended version of an earlier code written by \citet{Wolfe67}. The radial velocities are given a weight factor of 1.0 except for those obtained from four spectra that had a significantly lower S/N value (taken at HJD=55510.49097, 55545.38178, 55545.39625, 55545.40725). These radial velocities were given a weight factor of 0.5. To obtain accurate uncertainties, a wrapper with a Monte Carlo algorithm was written around the SBOP code. The input data were perturbed with the estimated uncertainties on the radial velocity measurements. The final uncertainties on the spectroscopic elements are determined from  5000 iterations in the Monte Carlo code.

As in the determination of the photometric elements, a circular orbit was assumed, fixing $e=0.0$ and $\omega = 90^{\circ}$. The resulting spectroscopic elements are given in Table \ref{tb_spectroscopic_elements}. The radial velocities of the components were analysed independently (single--lined spectroscopic binary SB1), but SB2 solutions lead to nearly identical results. 

\begin{table}
\centering
 \caption{Spectroscopic orbital solution for EF Aqr obtained with the SBOP code.}
\label{tb_spectroscopic_elements}
\begin{tabular}{lrr}
\hline\hline
Parameter	&	\multicolumn{1}{c}{Primary}	&	\multicolumn{1}{c}{Secondary}	\\\hline
$P$ (d)\tablefootmark{a}		&	\multicolumn{2}{c}{2.85357206}			\\
$T_0$\tablefootmark{a}			&	\multicolumn{2}{c}{2452556.57276}		\\
$e$\tablefootmark{a}			&	\multicolumn{2}{c}{0}				\\
$\omega$ ($^{\circ}$)\tablefootmark{a}	&	\multicolumn{2}{c}{90}				\\
$q$ 				&	\multicolumn{2}{c}{0.761$\pm$0.003}			\\
$\gamma$ (km s$^{-1}$)		&	$-$4.2341 $\pm$ 0.0694	&	$-$4.372 $\pm$ 0.0974	\\
$K$  (km s$^{-1}$)		&	84.2188 $\pm$ 0.0882	&	110.8169 $\pm$ 0.1238	\\
$a$ $\sin{i}$ (R$_{\odot}$)	&	3.3047 $\pm$ 0.0035	&	4.3484 $\pm$ 0.0049	\\
$M$ $\sin^3{i}$ (M$_{\odot}$)	&	1.244 $\pm$ 0.008	&	0.946 $\pm$ 0.006	\\
$\sigma$			&	0.28			&		0.39		\\
\hline
\end{tabular}
\tablefoot{\tablefoottext{a}{Assumed values.}\\
The quoted errors are formal errors from the iterative least--squares solution procedure.}
\end{table}

\begin{figure}
\centering
\includegraphics{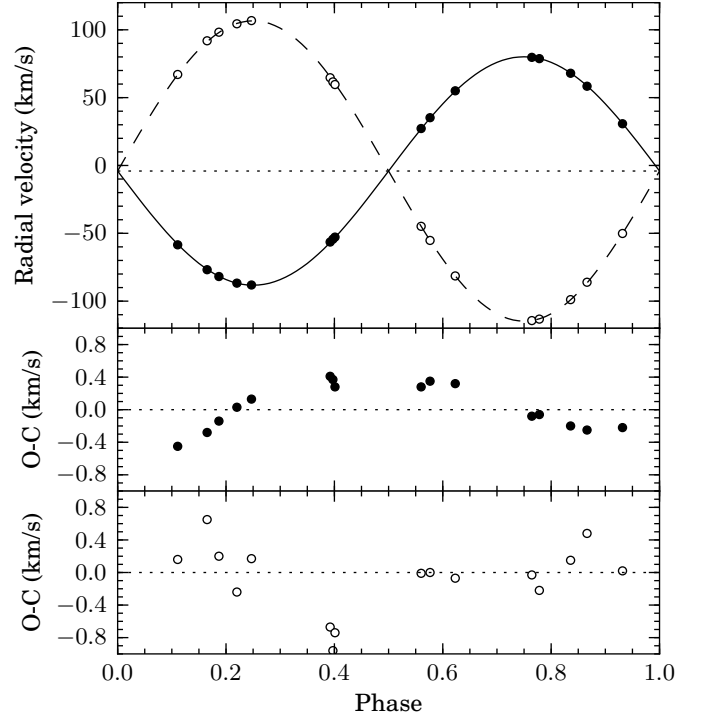}
\caption{Spectroscopic orbital solution for EF Aqr (solid line: primary, dashed line: secondary), and the observed radial velocities (filled circles: primary, open circles: secondary). Phase 0.0 corresponds to the central primary eclipse.}
\label{fig_rv_curves}
\end{figure}

In Fig \ref{fig_rv_curves} the spectroscopic orbital solution is plotted together with the residuals of both components. As can be seen from the plot, there is a clear sine--like pattern with an amplitude of around 500 m s$^{-1}$ in the residuals of both components. Including the eccentricity and argument of periastron as free parameters improves the solution, but the reduced masses and semi--major amplitudes do not change. In this way an eccentricity of $e=0.0018\pm0.0022$ and an argument of periastron of $\omega=90.23^{\circ}\pm0.21$ are found. However, the photometry (light curves and times of minima) limits the eccentricity to a maximum of $e=0.00005\pm0.00008$, indicating that the orbit is practically circular. The periodic residuals in the radial velocity curves can be explained by the activity of the system. The light curves show clear signs of star spots, and previous research of \cite{Huerta08} has shown that star spots can cause radial velocity variations of the order of 0.5 km s$^{-1}$. This theory can not be proven because there are no accurate photometric observations available that coincide with the spectroscopic observations. Altogether, these small deviations in the radial velocity curves do not have a significant effect on the resulting spectroscopic elements, and can be neglected.

\section{Chemical abundances}\label{chemical_abundances}
\begin{figure*}[!t]
\centering
\includegraphics{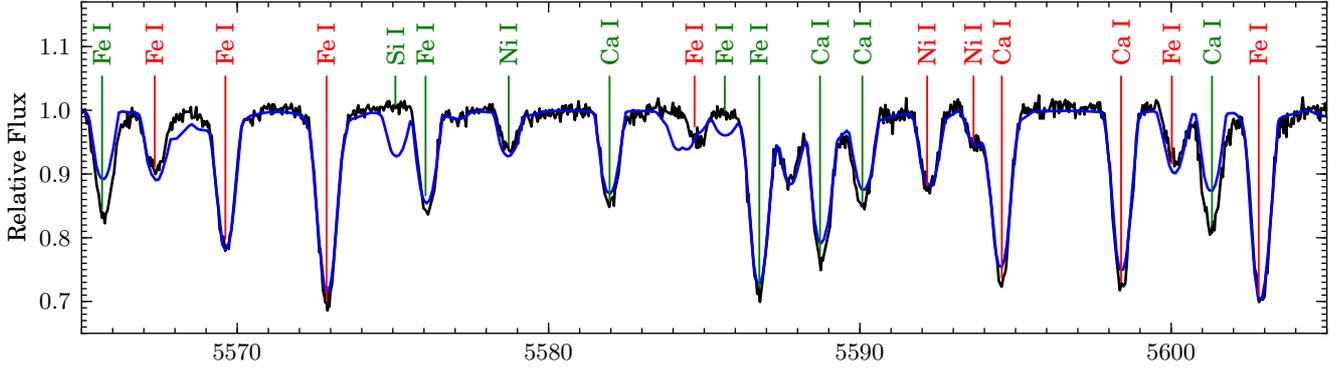}
\caption{40 \AA\ region centred at 5585 \AA\ of the disentangled spectrum for the primary component of EF Aqr (black) and a matching synthetic spectrum calculated with VWA (blue). The lines indicated in green are used in the abundance determinations.}\label{fig_spectrum}
\end{figure*}

To determine the chemical abundances of EF Aqr, we used all 17 HERMES spectra. Because sufficient spectra are available, the DisEntangler package of \citet{Simon94} was used to disentangle the spectra, so that the primary and secondary could be analysed separately. In this way a higher accuracy can be obtained than if composite spectra were used. The 35 orders used also for the radial velocity determinations, which had sufficient S/N and no influence of telluric lines, were disentangled individually for the abundance analysis. The DisEntrangler algorithm treats the spectra with an equal light ratio, resulting in component spectra that are determined independently of an additive constant that depends on the light ratio in each order. This constant is calculated using MARCS model atmospheres \citep{Gustafsson08}, and implemented in the normalisation process using dedicated IDL programs. DisEntangler also assumes a constant light level, but since EF Aqr is constant to within 0.5 \% outside eclipses, this is of no concern. A 40\AA\ region of the spectrum of the primary component, centred on 5585\AA\, is shown in Fig. \ref{fig_spectrum}.

The versatile wavelength analysis (VWA) tool of \citet{Bruntt02} was used to determine the abundance and temperature of both components of EF Aqr.  VWA used the SYNTH software of \citet{Valenti96} to generate synthetic spectra. Atmosphere models were interpolated from the MARCS model atmospheres \citep{Gustafsson08}, which use the solar composition of \citet{Grevesse07}. The atomic line data were taken from the VALD database \citep{Kupka99}. However, the $\log(gf)$ values were adjusted so that every line measured by \citet{Wallace98} reproduces the atmospheric abundances by \citet{Grevesse07}. For a detailed description of the VWA tool, the reader is referred to \citet{Bruntt04,Bruntt08,Bruntt09, Bruntt10a, Bruntt10b}.

When comparing the observed orders to synthetic spectra, the rotational velocities of the two components are found to be $v\sin{i}_p = 21\pm 4$ km s$^{-1}$ and $v\sin{i}_s = 18\pm 4$ km s$^{-1}$ respectively, indicating that they are fast rotators. Owing to the higher degree of line blending caused by the fast rotation, only 468 (primary) and 396 (secondary) lines were selected. The surface gravity was fixed at the values determined from the photometric and spectroscopic elements given in Section \ref{absolute_dimensions}. Spectroscopy alone is usually only capable to determine $\log{g}$ with an accuracy of up to 0.1 dex \citep{Bruntt10a, Valenti05}, but in a binary system accuracies as high as 0.01 dex can be obtained. The effective temperature and micro--turbulence were tuned to remove differences between {Fe \sc i} and {Fe \sc ii} abundance, and any correlation between abundance and excitation potential or equivalent width. In this way effective temperatures of $T_{\mathrm{eff,p}} = 6140 \pm 50$ K and $T_{\mathrm{eff,s}} = 5200 \pm 100$ K were found. \citet{Bruntt10a} found a systematic offset of 40 K between spectroscopicaly obtained temperatures and temperatures obtained with direct methods. After subtracting this offset spectroscopic temperatures of $T_{\mathrm{eff}} = 6100 \pm 50$ K (primary) and $T_{\mathrm{eff}} = 5160 \pm 100$ K (secondary) were obtained. 

The micro--turbulence is determined to be $\xi_p = 1.60 \pm 0.10$ km s$^{-1}$ and $\xi_s = 1.32 \pm 0.10$ km s$^{-1}$. For the primary this agrees very well with the predicted micro--turbulence of $\xi_p = 1.65 \pm 0.34$ km s$^{-1}$ by the calibration of \citet{Edvardsson93} (which is an analytical fit to models based on 157 stars). For the secondary the predicted micro--turbulence is $\xi_s = 0.66 \pm 0.38$ km s$^{-1}$, which is significantly lower than found in the abundance determination. However, the secondary's effective temperature is outside the range of the Edvardsson calibration, which is only valid in the temperature range $T_{\mathrm{eff}} = 5550 \rightarrow 6800 K$. \citet{Bruntt10a} derived a micro--turbulence relation using 23 solar--type stars. This calibration predicts $\xi = 1.27 \pm 0.13$ km s$^{-1}$ (primary) and $\xi = 0.84 \pm 0.11$ km s$^{-1}$ (secondary). Both the predictions for the primary and the secondary are significantly lower than determined in the present work from observations. We cannot pinpoint an exact reason for the difference between the calibrations and the results presented here, but it has to be noted that the implementation of the microturbulence can depend on the model atmosphere used, which could explain the difference between our results and the results of Edvardsson. 

The abundances determined from all useful lines with an equivalent width between 10 and 90 m\AA\  are presented in Table \ref{tb_abundances}. Obviously a very robust iron abundance of [Fe/H] = -0.002 $\pm$ 0.090 dex is obtained with excellent agreement between the results of {Fe \sc i} and {Fe \sc ii} lines of both components. Changing the temperature of the primary by $\pm50$ K and the secondary by $\pm100$ K results in a $\pm0.04$ dex change in {Fe \sc i} abundance and $\mp0.02$ dex change in {Fe \sc ii} abundance. If the micro--turbulence is increased or decreased by $\pm0.10$ km s$^{-1}$, the neutral iron abundance changes by $\pm0.07$ dex, while the ionised iron abundance changes by $\mp0.02$ dex. The abundances of the other ions listed in Table \ref{tb_abundances} agree well with the results obtained for the iron abundance and a relative abundance identical to the solar value. The few outliers such as {Cu \sc i}, {C \sc i}, {S \sc i}, and {Ce \sc ii} are based only on one line and can be neglected. The final metal abundance of EF Aqr is determined based on the abundances of all elements with eight or more accepted lines, and results in [M/H] = 0.00 $\pm$ 0.10 dex.

In addition to the spectroscopic abundance determinations, the calibration of \citet{Holmberg07} was used to determine the metalicity directly from the de-reddened $uvby$ indices for the individual components. This results in metallicities of [Fe/H] = $0.04 \pm 0.11$ (primary) and [Fe/H] = $-0.20 \pm 0.27$ (secondary), respectively. The photometric metallicity obtained for the primary corresponds to the spectroscopic metallicity within errors. For the secondary the difference is larger, but the two methods still agree within the estimated errors.

\begin{table}
\centering
\caption{The abundances ([El/H]) for the primary and secondary components of EF Aqr}
\label{tb_abundances}
\begin{tabular}{lrlrrlr}
\hline\hline
           & \multicolumn{3}{c}{Primary} & \multicolumn{3}{c}{Secondary}  \\
Ion	   	& 	[El/H] 	& 	rms 	& 	N\tablefootmark{a} 	& 	[El/H] 	& 	rms 	& 	N\tablefootmark{a} 	\\
\hline
{C  \sc   i}	&	$-$0.24	&	-	&	1	&		&		&		\\
{Na \sc   i}	&	0.08	&	-	&	1	&	0.09	&	-	&	1	\\
{Mg \sc   i}	&	0.01	&	-	&	1	&		&		&		\\
{Al \sc   i}	&	0.17	&	-	&	1	&		&		&		\\
{Si \sc   i}	&	0.06	&	0.08	&	10	&	0.10	&	0.12	&	3	\\
{Si \sc  ii}	&	0.07	&	0.09	&	2	&		&		&		\\
{S  \sc   i}	&	0.35	&	-	&	1	&		&		&		\\
{Ca \sc   i}	&	0.01	&	0.09	&	5	&	0.00	&	0.19	&	6	\\
{Sc \sc  ii}	&	$-$0.09	&	0.11	&	2	&	0.12	&	0.17	&	2	\\
{Ti \sc   i}	&	$-$0.04	&  	0.10	&	2	&	$-$0.09	&	0.15	&	3	\\
{Ti \sc  ii}	&	$-$0.06	&	0.09	&	3	&	$-$0.03	&	-	&	1	\\
{V  \sc   i}	&	$-$0.15	&  	0.11	&	2	&	0.24	&	0.11	&	8	\\
{Cr \sc   i}	&	$-$0.09	&	0.11	&	9	&	0.10	&	0.11	&	4	\\
{Cr \sc  ii}	&	$-$0.02	&	0.08	&	4	&		&		&		\\
{Mn \sc   i}	&	$-$0.06	&	0.12	&	4	&	0.18	&	0.26	&	4	\\
{Fe \sc   i}	&	0.00	&	0.08	&	136	&	0.00	&	0.10	&	88	\\
{Fe \sc  ii}	&	0.00	&	0.09	&	11	&	0.00	&	-	&	1	\\
{Co \sc   i}	&	0.13	&  	0.12	&	2	&	$-$0.06	&  	0.17	&	2	\\
{Ni \sc   i}	&	$-$0.06	&	0.08	&	18	&	0.12	&	0.15	&	10	\\
{Cu \sc   i}	&	$-$0.42	&	-	&	1	&	$-$0.62	&	-	&	1	\\
{Zn \sc   i}	&	$-$0.19	&	-	&	1	&		&		&		\\
{Y  \sc  ii}	&	$-$0.06	&	-	&	1	&	0.41	&	-	&	1	\\
{Ce \sc  ii}	&	$-$0.59	&	-	&	1	&		&		&		\\
\hline
\end{tabular}
\tablefoot{\tablefoottext{a}{Number of lines used per ion, the rms error is computed if two or more lines are available.}}
\end{table}

\section{Absolute dimensions}\label{absolute_dimensions}
The absolute dimensions of EF Aqr are presented in Table \ref{tb_absolute_dimensions}, as calculated from the photometric and spectroscopic elements given in Tables \ref{tb_adopted_photometric_elements} and \ref{tb_spectroscopic_elements}. The accuracy for the masses is better then one percent while for the radii an accuracy of 1.2\% was established. The $V$ magnitudes and $uvby$ indices were calculated from the combined magnitudes and indices of the system outside the eclipses (see Table \ref{tb_photometric_data}). The $V$ magnitude and $uvby$ indices obtained for the primary component during secondary eclipse agree reasonably well with those obtained from the photometric solution. The $E(b-y)$ interstellar reddening was calculated using the calibration of \citet{Olsen88}, using the $ubvy\beta$ standard photometry for combined light outside eclipses. 

The temperature of both components was determined using the three different colour-metallicity-temperature calibrations of \citet{Alonso96}, \citet{Holmberg07} and \citet{Casagrande10}, the spectroscopic derivation of \citet{Popper80}, and for the secondary also the flux scale method of this author. The results for each method are shown in Table \ref{tb_temperatures}. The uncertainties include the error on the colour indices, the metallicity, reddening and the uncertainty of the calibrations themselves. The obtained temperatures are mutually consistent within the errors of the calibrations, and as final temperatures the average values over the different methods are adopted: $T_{eff,p}=6150 \pm 65$ K and $T_{eff,s}=5185 \pm 110$ K.

The projected rotational velocities from the broadening functions  $v\sin{i}_p = 23$ km s$^{-1}$ and $v\sin{i}_s = 17$ km s$^{-1}$ agree with those from the abundance analysis in the previous section.
The projected rotational velocities agree within the errors with the calculated velocities assuming a synchronous circular orbit: $v\sin{i}_p = 23.7$ km s$^{-1}$ and $v\sin{i}_s = 16.9$ km s$^{-1}$. Furthermore, the age  ($1.5\pm0.6$ Gyr) determined in Section \ref{Discussion} is clearly older than the circularisation and synchronisation time scale as determined by \citet{Zahn77}: $\tau_{circ} = 0.28$ Gyr and $\tau_{syn} = 0.09$ Gyr, and by Claret \citep[and references herein]{Torres09}: $\tau_{circ} = 0.42$ Gyr, $\tau_{syn}(\textrm{prim}) = 0.21$ and $\tau_{syn}(\textrm{sec}) = 0.20$.

The distance to the system was calculated using the classical relation (see e.g. \citetalias{Clausen08}), adopting the solar values and bolometric corrections given in Table \ref{tb_absolute_dimensions}. For the primary a distance of $170.3 \pm 4.5$ pc was found which corresponds well with the distance of $174.4 \pm 6.3$ pc determined for the secondary component. For the system distance the average of both was adopted at $172.0 \pm 3.8$ pc.

\begin{table}
\centering
\caption{Astrophysical data for EF Aqr}\label{tb_absolute_dimensions}
 \begin{tabular}{lrr}
\hline\hline
\multicolumn{3}{l}{Systemic parameters}	\\
$P$ (d)		&	\multicolumn{2}{c}{2.85357206}			\\
$T_0$ (HJD)	&	\multicolumn{2}{c}{52556.5728}			\\
$e$		&	\multicolumn{2}{c}{0.0}				\\
$\omega$	&	\multicolumn{2}{c}{90.0}			\\
$a$ (R$_{\odot}$)	&	\multicolumn{2}{c}{10.9984 $\pm$ 0.0207}	\\
$v_0$ (km s$^{-1}$)	&	\multicolumn{2}{c}{$-$4.0660 $\pm$ 0.1264}	\\
$q$		&	\multicolumn{2}{c}{0.761 $\pm$ 0.003}		\\
$i$ $(^o)$	&	\multicolumn{2}{c}{88.454 $\pm$ 0.076}		\\
$E(b-y)$		&	\multicolumn{2}{c}{0.018 $\pm$ 0.011}		\\
$[$Fe/H$]$ (dex)	&	\multicolumn{2}{c}{0.00 $\pm$ 0.10}		\\
$d$ (pc)	&	\multicolumn{2}{c}{172.0 $\pm$ 3.8}		\\
\\
\multicolumn{3}{l}{Component parameters}	\\
	&	\multicolumn{1}{c}{Primary}	&	\multicolumn{1}{c}{Secondary}	\\
$M$ (M$_{\odot})$	&	1.244	$\pm$	0.008	&	0.946	$\pm$	0.006	\\
$R$ (R$_{\odot})$	&	1.338	$\pm$	0.012	&	0.956	$\pm$	0.012	\\
$K$ (km s$^{-1})$		&	84.180 $\pm$ 0.185	&	110.686 $\pm$ 0.302	\\
$v\sin{i}$ (km s$^{-1})$	&	21 $\pm$ 4		&	18 $\pm$ 4		\\
$\xi$ (km s$^{-1})$		&	1.60 $\pm$ 0.10		&	1.32 $\pm$ 0.10		\\
$\log{g}$ (cgs)	&	4.280	$\pm$	0.007	&	4.453	$\pm$	0.011	\\
$V_0$			&	10.015	$\pm$	0.022	&	11.711	$\pm$	0.024	\\
$(b-y)$			&	0.338	$\pm$	0.010	&	0.484	$\pm$	0.016	\\
$m_1$			&	0.183	$\pm$	0.010	&	0.332	$\pm$	0.021	\\
$c_1$			&	0.383	$\pm$	0.015	&	0.292	$\pm$	0.029	\\
$T_{\textrm{eff}}$ (K)		&	6150	$\pm$	65	&	5185	$\pm$	110	\\
$M_{\textrm{bol}}$ 		&	3.847	$\pm$	0.029	&	5.319	$\pm$	0.051	\\
$\log{L/L_{\odot}}$	&	0.360	$\pm$	0.011	&	$-$0.227	$\pm$	0.020	\\
BC 			&	$-$0.021 $\pm$ 0.013	&	$-$0.192 	$\pm$ 0.035	\\
$M_V$ 			&	3.868	$\pm$	0.030	&	5.511	$\pm$	0.0523	\\
\hline
 \end{tabular}
\tablefoot{Bolometric corrections (BC) by \citet{Flower96} were assumed together with $T_{\textrm{eff},\odot} = 5780$ K, $BC_{\odot}=-0.08$ and $M_{\textrm{bol},\odot} = 4.74$.}
\end{table}

\begin{table}
\centering
\caption{The effective temperature determined with five different methods.}\label{tb_temperatures}
\begin{tabular}{lr@{$\pm$}lr@{$\pm$}l}
\hline\hline
Method & \multicolumn{2}{c}{Primary}& \multicolumn{2}{r}{Secondary}\\ \hline
Alonso calibration\tablefootmark{a} 		& 6128 & 60 &\ \  5205 & 86\\
Holmberg calibration\tablefootmark{b} 		& 6165 & 68 &\ \  5248 & 109\\
Casagrande calibration\tablefootmark{c} 		& 6244 & 85 &\ \  5301 & 151\\
Popper flux scale\tablefootmark{d} 		& \multicolumn{2}{c}{/}&\  \ 5098 & 58\\
Spectroscopic temperature 	& 6100 & 50 &\ \  5160 & 103\\
\hline
\end{tabular}
\tablebib{
\tablefoottext{a}{\citet{Alonso96}}
\tablefoottext{b}{\citet{Holmberg07}}
\tablefoottext{c}{\citet{Casagrande10}}
\tablefoottext{d}{\citet{Popper80}}
}
\end{table}

\section{Stellar activity}
\begin{figure}
\centering
\includegraphics{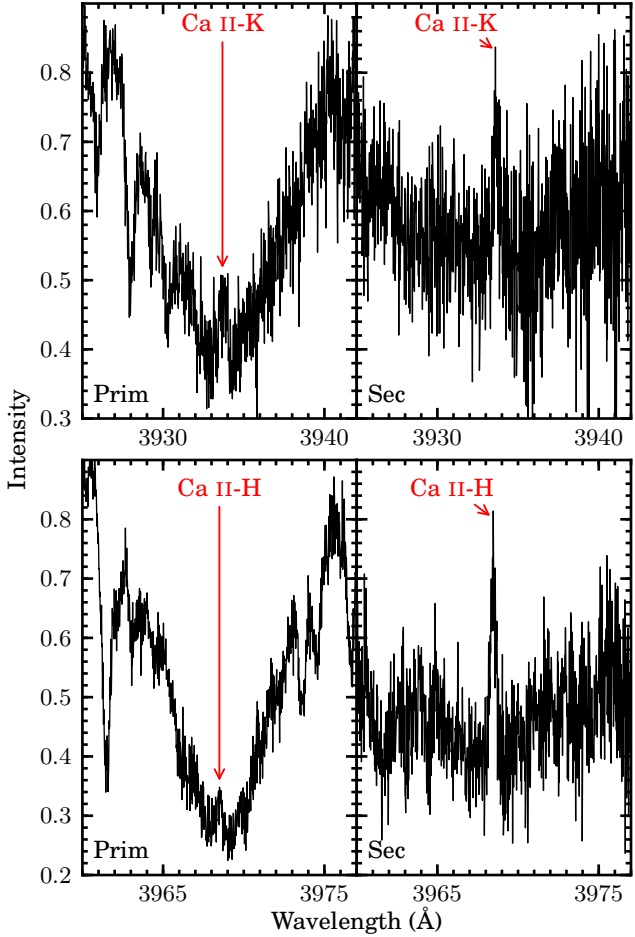}
\caption{{Ca \sc ii}-K (3933.68 \AA) and {Ca  \sc   ii} - H (3968.49 \AA) emission lines in the spectrum of both components.}
\label{fig_ca_emisson}
\end{figure}

The light curves of EF Aqr show periodic variations that could be caused by star spots. The period of the variations differs, and the anomalies in the light curves disappear and reappear during the eight years over which the observations were made. Several subsets of the lightcurves were analysed separately, but it proved impossible to fit the variations. However, they are a clear indicator of the activity of one or both components. The Rossby numbers, defined as the ratio of the rotation period over the convective turnover time, for the primary and secondary components are approximately 0.71 and 0.08. The limit for higher magnetic activity due to stronger dynamo action was determined at 0.65 by \citet{Hall94}. This places the secondary well within the active region, and the primary at the border. The most compelling evidence for activity of both components is the {Ca  \sc   ii} emission. In Fig \ref{fig_ca_emisson} the {Ca  \sc   ii} - K (3933.68 \AA) and {Ca  \sc   ii} - H (3968.49 \AA) emission lines for both components are shown. In the spectrum of the primary, the emission lines appear in the hydrogen absorption lines, while for the secondary, the {Ca \sc ii} lines are visible as emission stronger than the continuum. These signs show that the primary is slightly active, while the secondary is very likely to exhibit strong chromospheric activity and star spots.

Although the primary is less active than the secondary, it is prossible that starsports are the cause of the periodic variations in the radial velocity curve. Both components are suffuciently different in temperature, mass and radius that the effect of starspots can differ strongly for both components. The primary being the larger and hotter component is likely to exhibit a stronger effect from starspots.

\section{Discussion}\label{Discussion}
In this section the absolute dimensions obtained for EF Aqr are compared with the properties of recent theoretical stellar evolutionary models. For an extended description of the models, the reader is referred to \citetalias{Clausen08}.

\subsection{Comparison with solar-calibrated models}
\begin{figure}
\centering
\includegraphics{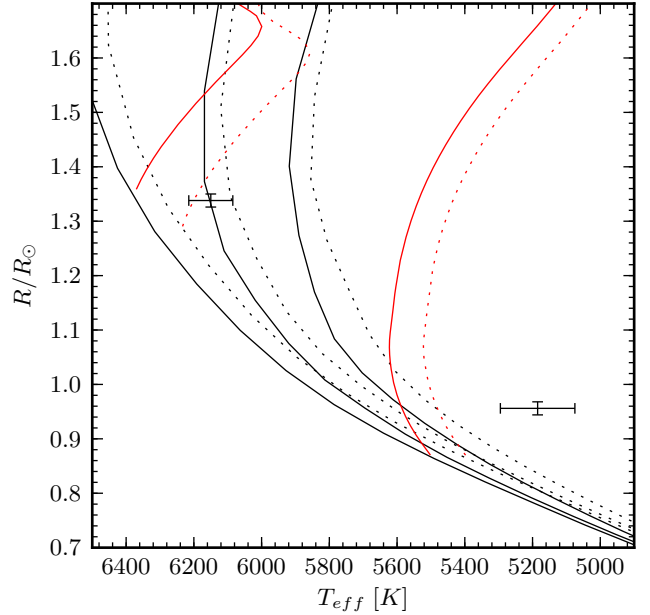}
\caption{EF Aqr compared with $Y^2$ models for [Fe/H] = 0.00 (solid line) and [Fe/H] = +0.10 (dotted line). Isochrones for 2.25, 3.5 and 6.5 Gyr (left to right) are shown in black and tracks for the component masses are shown in red.}
\label{fig_yy_T_R}
\end{figure}

\begin{figure}
\centering
\includegraphics{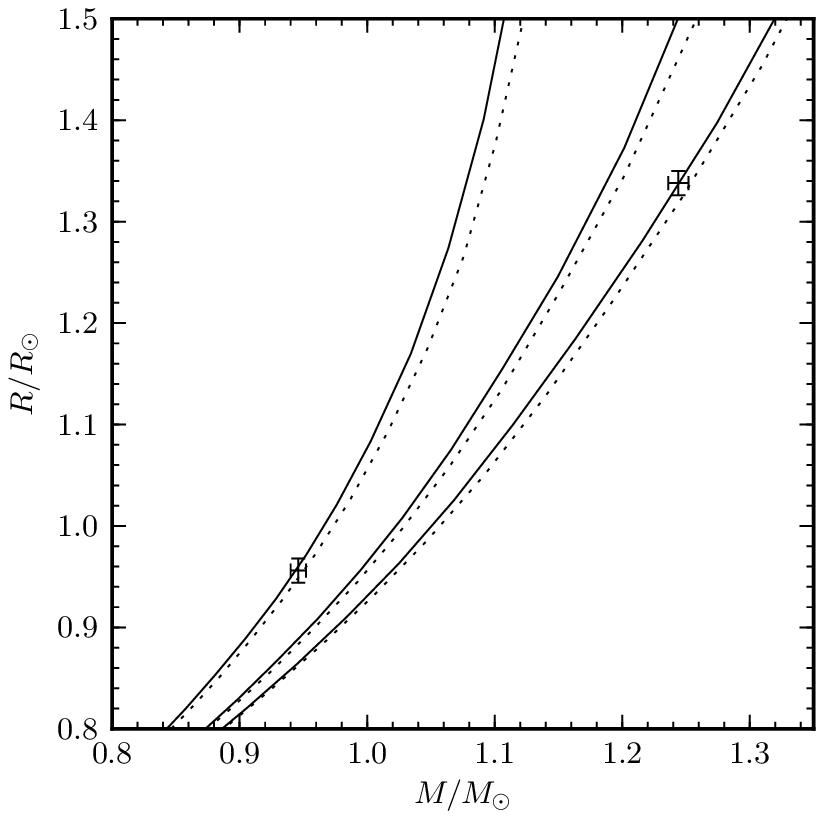}
\caption{EF Aqr compared with $Y^2$ models for [Fe/H] = 0.00 (solid line) and [Fe/H] = +0.10 (dotted line). Isochrones for 2.25, 3.5 and 6.5 Gyr are shown.}
\label{fig_yy_M_R}
\end{figure}

\begin{figure}
\centering
\includegraphics{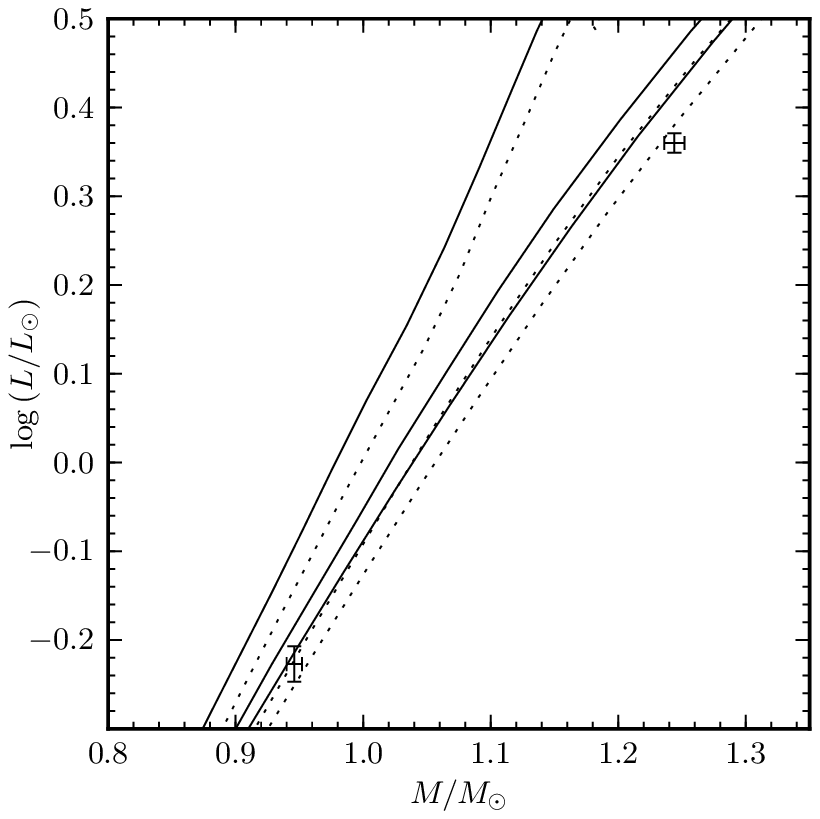}
\caption{EF Aqr compared with $Y^2$ models for [Fe/H] = 0.00 (solid line) and [Fe/H] = +0.10 (dotted line). Isochrones for 2.25, 3.5 and 6.5 Gyr are shown.}
\label{fig_yy_M_L}
\end{figure}

Figures \ref{fig_yy_T_R}--\ref{fig_yy_M_L} illustrate the comparison with the Yonsei-Yale ($Y^2$) evolutionary tracks and with the isochrones by \citet{Demarque04}. The mixing length parameter in the convective envelopes was calibrated using the Sun, and was held fixed at $l/H_p=1.7432$. An enrichment law of $Y=0.23+2Z$ was adopted, together with the solar mixture of \citet{Grevesse96}, leading to $(X,Y,Z)_{\odot}=(0.71564,0.26624,0.01812)$. Only models without $\alpha$-element enrichment were considered ([$\alpha$/Fe]=0.0). 

Fig. \ref{fig_yy_T_R} shows that theoretical tracks for the observed masses and metalicity are hotter than the observations. Especially for the secondary, the temperature difference is large ($\sim400$ K). On the other hand, the theoretical temperature difference between the components of 800 K is only slightly smaller than the observed temperature difference of $965 \pm 125$ K, which is partly covered by the uncertainty on the track positions caused by the small mass uncertainty. To incorporate the $0.10$ dex uncertainty on the metalicity, theoretical tracks with [Fe/H] = +0.10 were compared to the observed parameters, too. As was shown in Sec. \ref{chemical_abundances}, a higher metallicity gives rise to a higher spectroscopic temperature, leading to an increase of $\sim20$ and $\sim40$ K for the primary and secondary component. This higher metalicity track fits the observed temperature and radius of the primary within errors. For the secondary component, the theoretical parameters correspond better to the observed results, but the theoretical temperature is still 250 K hotter than observed, which is well beyond the uncertainty. 

The masses and radii calculated from the light and radial velocity curves are independent of the distance to the system (scale--independent). The mass--radius plane plotted in Fig. \ref{fig_yy_M_R} shows that the model predicts an age difference of 4 Gyr between the components, 2.25 Gyr for the primary and 6.5 Gyr for the secondary. Changing to isochrones with a higher metallicity makes no significant difference in this result. At the age of the primary, the predicted radius of the secondary would be 0.869 R$_{\odot}$, which is 9 \% smaller than observed, while the observations have a precision of $\sim$1\%. The mass-luminosity relation, shown in Fig \ref{fig_yy_M_L}, predicts an age of around 2 Gyr for both components, with a metallicity of [Fe/H]=+0.10, or a younger age of around 1 Gyr with a metallicity of [Fe/H]=0.0. 

The same picture is seen in the Victoria-Regina \citep{Vandenberg06} and BaSTI \citep{Pietrinferni04} models, which differ only slightly from the $Y^2$ models with respect to the input physics and He enrichment laws. A short description of these models is given in \citetalias{Clausen08}.

In conclusion, the secondary component of EF Aqr is significantly larger, and about 400 K cooler than predicted by models, which adopt a mixing length parameter matching that of the Sun. The primary component agrees better with the models. Both components, however, reveal a serious discrepancy between models and the observations, which we attribute to problems in the modelling, since the discrepancies are much larger than the observational uncertainty, which is seen particularly clearly in Figs.\ref{fig_yy_T_R} and \ref{fig_yy_M_R}. 

\subsection{Comparison with mixing-length tuned models}
Several authors have demonstrated that models that adopt a reduced mixing length parameter are better at fitting active solar--type binary components. Both components of EF Aqr show clear signs of increased magnetic activity (in particular the secondary component), and have relatively high rotational velocities ($>$ 10 km s$^{-1}$ see e.g. \citetalias{Clausen09}), making EF\,Aqr a good candidate to compare to mixing-length tuned models.

For this purpose, dedicated models for EF Aqr were computed with the GRANADA code by \citet{Claret04} 
, which assumes an enrichment law of Y = 0.24 + 2.3Z together with the solar mixture by \citet{Grevesse98}, leading to (X,Y,Z)$_{\odot}$=(0.704,0.279,0.017). In this model, the mixing length parameter needed to reproduce the Sun is $l/H_p=1.68$.

\begin{figure}
\centering
\includegraphics{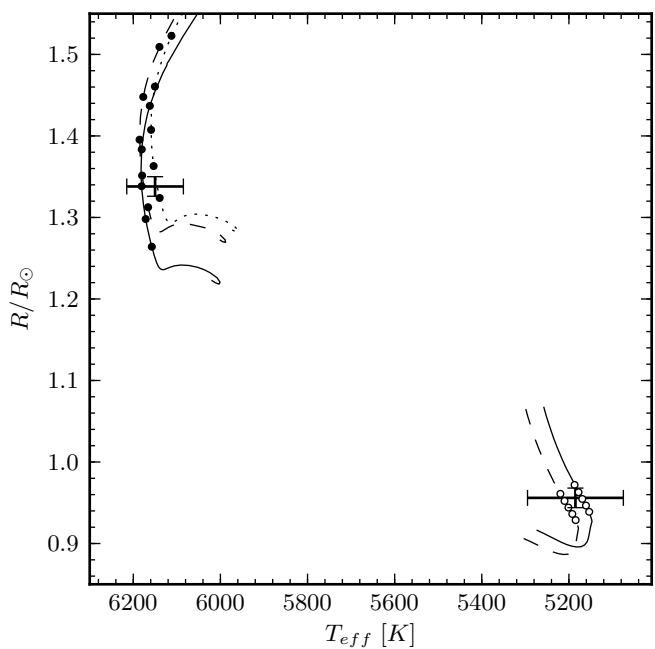}
\caption{EF Aqr compared with Granada models for the observed masses calculated with different mixing length parameters (see Table \ref{tb_granada}). Primary component (filled cicles, upper left part of diagram): Case 1: full line, case 2: dashed line, case 3: dotted line. Secondary component (open circles, lower right part of diagram): case 4: full line and case 5: dashed line. The circles on the evolutionary tracks mark ages of 0.5, 1, 1.5, 2, and 2.5 Gyr.}
\label{fig_granada_tracks}
\end{figure}

\begin{table}
\centering
\caption{Ages of the primary and secondary components of EF Aqr determined from a comparison between the observed masses and radii with the Granada models for different mixing length parameters.}\label{tb_granada}
\begin{tabular}{cccc}
\hline\hline
  & Type & $l/H_p$ & Age\\\hline
1 & P	& 1.30	& $1.50\pm0.18$ \\
2 & P	& 1.20	& $0.90\pm0.20$ \\
3 & P	& 1.15	& $0.70\pm0.20$ \\
4 & S	& 1.05	& $1.55\pm0.60$ \\
5 & S	& 1.10	& $2.25\pm0.75$ \\
\hline
\end{tabular}
\end{table}

Several tracks for the precise observed masses were calculated. Tests on the influence of the helium content were performed, and the helium content stated above gave the best results. The five best tracks (three for the primary and two for the secondary) are listed with there resulting ages in Table \ref{tb_granada}.
The temperature-radius comparison between EF Aqr and the Granada models is shown in Fig. \ref{fig_granada_tracks}. 
The primary component is best fitted with the $l/H_p=1.15$ track at an age of $0.70\pm0.20$ Gyr, while the best fit for the secondary component is obtained with an $l/H_p=1.05$ track with resulting age of $1.55\pm0.60$ Gyr. 
When a slightly longer mixing length of $l/H_p=1.30$ is used, the age of the primary can be fitted at $1.50\pm0.18$ Gyr. The correspondence with the temperature is slightly poorer, but still well within the uncertainty. Granada models with a solar mixing length parameter are unable to fit both components. 

The best--fitting tracks with identical derived ages for the two components are those with mixing length $l/H_p=1.30$ for the primary, and $l/H_p=1.05$ for the secondary. These tracks predict a common age of $1.5\pm0.6$ Gyr for the system. This does imply that the mixing length parameter for both components can be tuned with reasonable precision, provided that accurate dimensions and abundances are available. 

\subsection{Comparison with other binaries}
An overview of the properties of 11 solar--type eclipsing binaries of which the masses and radii are determined with a precision of at least 2\% is given in \citetalias{Clausen09}. This list can be extended with EW Orionis \citep{Clausen10} and IM Virginis \citep{Morales09}. Clausen divided these binaries into two groups, the first consisting of active binaries with high rotational velocities, and the second group with longer periods and no signs of activity. (IM Vir belongs to the first group, while EW Ori belongs to the second). The non--active binaries with a longer period can be fitted by solar--calibrated models, while the active binaries can be fitted with mixing--length--tuned models. EF Aqr is then a clear member of this second group, and supports the suggestion of \citet{Clausen10} that the mixing length parameter is governed by two main effects: For inactive main sequence stars a slight decrease in mixing length with increasing temperature and mass, and secondly, for active stars a strong decrease in mixing length compared with inactive stars of the same mass.

\section{Summary and conclusions}
From state-of-the-art observations and analyses, precise (0.6-1.0\%) absolute dimensions have been obtained for the totally eclipsing G0 system EF Aqr. A detailed spectroscopic analysis yielded a Fe abundance relative to the Sun of [Fe/H]=0.00$\pm$0.10, with the abundances of Si, Ca, Sc, Ti, V, Cr, and Co and Ni being close to solar, too.

The 0.956 M$_{\odot}$ secondary is found to be active with star spots and strong {Ca \sc ii} H and K emission. The 1.244 M$_{\odot}$ primary shows signs of activity as well, but at a lower level. The system orbit is circular, and the measured rotational velocities, 21$\pm$4 km s$^{-1}$ (primary) and 18$\pm$4 km s$^{-1}$ (secondary) excellently agreement with theoretically predicted rotational velocities for synchronous rotation.

We showed that theoretical stellar models with solar-value mixing length parameters (Yonsei-Yale, Victoria-Regina and BaSTI) are unable to fit both components of EF Aqr at a common age. At the derived age of the primary, the secondary was found to be 9\% larger than predicted. Furthermore, the components were observed to be about 200 K (primary) and 400 K (secondary), cooler than predicted by the models.

Models that adopt a significantly lower mixing length parameter can remove these discrepancies. Granada models calculated for the observed metalicity with a mixing length of $l/H_p=1.30$ (primary) and $l/H_p=1.05$ (secondary) reproduce the properties of both components of EF Aqr at a common age of 1.5$\pm$0.6 Gyr.

EF Aqr falls in the group of solar--type systems with high rotational velocities that cannot be matched by solar-calibrated theoretical models as defined by \citetalias{Clausen09}. It supports the suggestion that magnetic activity is the likely cause for the radius and temperature discrepancies, which means in turn that active binaries have a significantly lower mixing length than their inactive counterparts of similar mass.

This study is part of a larger project on solar-type eclipsing binaries: see \citetalias{Clausen01}

\begin{acknowledgements}
      In memory of Jens Viggo Clausen, who passed away on June 6, 2011.
      It is a great pleasure to thank the following colleagues and students, who have shown interest in our project and have participated in the (semi)automatic observations of EF Aqr at the SAT: Sylvain Bouley, Christian Coutures, Thomas H. Dall, Pascal Fourqu\'e, Anders Johansen, Erling Johnsen, Raslan Leguet, Alain Maury, Christina Papadaki, John D. Pritchard, Samuel Regandell, and Chris Sterken. Excellent technical support was received from the staffs of Copenhagen University and ESO, La Silla. 
      The spectroscopic observations were obtained with the HERMES spectrograph, which is supported by the Fund for Scientific Research of Flanders (FWO), Belgium , the Research Council of K.U.Leuven, Belgium, the Fonds National Recherches Scientific (FNRS), Belgium, the Royal Observatory of Belgium, the Observatoire de Genève, Switzerland and the Thüringer Landessternwarte Tautenburg, Germany.
      The projects ``Stellar structure and evolution - new challenges from ground and space observations'' and ``Stars: Central engines of the evolution of the Universe'', carried out at Copenhagen University and Aarhus University, are supported by the Danish National Science Research Council (FNU).
      The following internet-based resources were used in research for this paper: the NASA Astrophysics Data System; the SIMBAD database and the VizieR service operated by CDS, Strasbourg, France; the ar$\chi$ive scientific paper preprint service operated by Cornell University.
      This publication makes use of data products from the Two Micron All Sky Survey, which is a joint project of the University of Massachusetts and the Infrared Processing and Analysis Center/California Institute of Technology, funded by the National Aeronautics and Space Administration and the National Science Foundation. 
      The research leading to these results has received funding from the European Research Council under the European Community's Seventh Framework Programme (FP7/2007--2013)/ERC grant agreement N$^{\underline{\mathrm o}}$\,227224 ({\sc prosperity}), as well as from the Research Council of K.U.Leuven grant agreement GOA/2008/04.
\end{acknowledgements}

\bibliographystyle{aa}
\bibliography{references}

\end{document}